\documentclass[a4paper,11pt]{article}
\usepackage{pos}
\usepackage{colortbl}
\usepackage{tabularx}

\definecolor{EPPS21color}{RGB}{68,119,170} 
\definecolor{nNNPDF30color}{RGB}{34,136,51} 
\definecolor{nCTEQ15WZSIHcolor}{RGB}{170,51,119} 
\definecolor{TUJU21color}{RGB}{204,187,68} 
\definecolor{KSASG20color}{RGB}{102,204,238} 

\title{Nuclear PDFs at the beginning of LHC Run 3}

\author*[a,b]{Petja Paakkinen}

\affiliation[a]{University of Jyväskylä, Department of Physics,\\
  P.O. Box 35, FI-40014 University of Jyväskylä, Finland}

\affiliation[b]{Helsinki Institute of Physics,\\
  P.O. Box 64, FI-00014 University of Helsinki, Finland}

\emailAdd{petja.k.m.paakkinen@jyu.fi}

\abstract{Nuclear parton distribution functions (nuclear PDFs), collinearly factorizable in perturbative QCD (pQCD), are currently extracted most reliably from global analyses of experimental data. Recently, the progress in this field has been driven mainly by the new data from LHC proton-lead collisions. I will review the progress in using data from LHC Run 1 and 2 and also hint at some opportunities and challenges that we face during LHC Run 3.}

\FullConference{%
  The Tenth Annual Conference on Large Hadron Collider Physics - LHCP2022\\
  16-20 May 2022\\
  online
}

\renewcommand{\logo}{\relax}

\begin{document}
\maketitle

\vspace{-0.275cm}
\section{Introduction}
\vspace{-0.175cm}

The field of extracting collinearly factorisable nuclear parton distribution functions (nPDFs) through global analyses of experimental data has been very active in the past years.
Table~\ref{Tbl:nPDFs} sum\-maris\-es the inputs and methodological choices in recent nPDF analyses~\cite{Khanpour:2020zyu,Helenius:2021tof,Eskola:2021nhw,AbdulKhalek:2022fyi,Duwentaster:2022kpv}.
Though the impact of the new deep-inelastic scattering (DIS) data from JLab on large-$x$ nPDFs has also been recently studied~\cite{Paukkunen:2020rnb,Segarra:2020gtj}, and the role of neutrino-nucleus ($\nu$A) DIS data in nuclear PDFs is still under debate~\cite{Muzakka:2022wey}, the progress in this field has been driven mainly by the inclusion of new data from LHC proton-lead (pPb) collisions, including the production of electroweak (EW) bosons, high-$p_{\rm T}$ jets and heavy quarks (HQ). These processes and their impact on the extraction of nuclear gluon distributions will be the main topic in this proceedings contribution.

\vspace{-0.275cm}
\section{EW bosons and high-$p_{\rm T}$ jets}
\vspace{-0.175cm}

Evident from Table~\ref{Tbl:nPDFs}, almost every recent nPDF analysis includes a set of data from EW boson production in pPb collisions at the LHC. These observables are an important probe of the flavour separation of nuclear modifications, but carry also sensitivity to nuclear gluons since small-$x$, high-$Q^2$ quarks and gluons are correlated through DGLAP evolution. This includes the most precise CMS measurement of $W^\pm$ production at 8.16 TeV~\cite{CMS:2019leu}. A possible complication in interpreting these data arises from the fact that the published absolute cross sections are sensitive also to proton-PDF uncertainties. For this reason, EPPS21~\cite{Eskola:2021nhw} constructs nuclear modification ratios of the data to get more direct access to the nuclear modifications of the PDFs, whereas TUJU21~\cite{Helenius:2021tof}, nNNPDF3.0~\cite{AbdulKhalek:2022fyi} and nCTEQ15HQ~\cite{Duwentaster:2022kpv} choose to fit to the absolute cross sections. Of these analyses, EPPS21 and nNNPDF3.0 also propagate the uncertainties from the free-proton PDFs in the fit. While this gives a subleading effect when nuclear ratios are used~\cite{Eskola:2021nhw}, it is possible that with the increased data precision at Run~3, the proton-PDF uncertainties can begin to have a larger impact in the nPDF fits~\cite{Eskola:2022rlm}. The new 8.16 TeV Drell-Yan (DY) data from CMS~\cite{CMS:2021ynu} have also been studied in some of the most recent analyses, with the data in the lower mass bin showing first direct evidence for the need to incorporate NNLO corrections in the nPDF analyses~\cite{Helenius:2021tof,AbdulKhalek:2022fyi}, but also the higher mass bin exhibiting large fluctuations that are causing some tension between the data and the nPDF fits~\cite{Helenius:2021tof,Eskola:2021nhw,AbdulKhalek:2022fyi}.

The EPPS21 and nNNPDF3.0 analyses include also the double-differential dijet data from 5.02 TeV pPb~\cite{CMS:2018jpl} using the nuclear ratios of self-normalised spectra, where theoretical uncertainties from hadronisation and free-proton baseline are expected to cancel, thus giving direct access to the PDF nuclear modifications. While otherwise finding a good fit, both analyses report on a difficulty in describing the most forward data points. Even though the scale uncertainties have been shown to cancel very effectively in the nuclear ratios~\cite{Eskola:2019dui}, it is not ruled out that NNLO corrections could improve the description also here through changes in the parton evolution and sensitivities to different channels and $x$ ranges. Also larger-than-expected non-perturbative corrections could in principle modify the nuclear ratios. Repeating the measurement with Run~2 and 3 data could shed more light on the issue.

\vspace{-0.275cm}
\section{Different approaches for fitting HQ data}
\vspace{-0.175cm}

As indicated in Table~\ref{Tbl:nPDFs}, the approaches in fitting the HQ data vary between analyses. For example, EPPS21 uses the S-ACOT-$m_{\rm T}$ general-mass variable-flavour-number (GMVFN) scheme~\cite{Helenius:2018uul}, while nNNPDF3.0 relies on fixed-order+parton-shower (FO+PS) predictions from POWHEG+PYTHIA. The latter falls into the category of fixed-flavour-number scheme (supplemented with parton shower and hadronisation from a Monte-Carlo event generator), where the heavy quarks are produced only at the matrix element level, whereas in the GMVFN scheme these are produced also in the initial and final state radiation. Importantly, the GMVFN calculation includes gluon-to-HQ fragmentation, which can have a large contribution to the cross section~\cite{Helenius:2018uul} and thus alter the small-$x$ sensitivity of the observable~\cite{Eskola:2019bgf}. Interestingly, nNNPDF3.0, with their POWHEG+PYTHIA approach, finds a large scale uncertainty in the D$^0$-production nuclear ratios and therefore choose to include only the more constraining forward data. In the GMVFN scheme, these scale uncertainties instead very effectively cancel in the nuclear ratios for $p_{\rm T} > 3$~GeV~\cite{Eskola:2019bgf}.

\begin{table}
  \small
  \centering
  \newcolumntype{Y}{>{\centering\arraybackslash}X}
  \renewcommand{\arraystretch}{0.795}
  \begin{tabularx}{3.0cm}{|Y|}
    \hline
    {\footnotesize \phantom{$^\dag$nPDF$^\dag$}} \\
    \hline
    \hline
    Order in $\alpha_s$ \\
    $l$A NC DIS \\
    $\nu$A CC DIS \\
    pA DY \\
    $\pi$A DY \\
    RHIC\,dAu\,{\footnotesize $\pi^0,\!\pi^\pm$} \\
    LHC\,pPb\,{\footnotesize $\pi^0,\!\pi^\pm,\!K^\pm$} \\
    LHC\;pPb\;dijets \\
    LHC\;pPb\;HQ \\
    LHC\;pPb\;W,Z \\
    LHC\;pPb\;{\footnotesize dir.-}$\gamma$ \\
    \\
    $Q, W$ cut in DIS \\
    $p_{\rm T}$ cut in {\footnotesize HQ,$\pi$,$K$} \\
    Data points \\
    Free parameters \\
    Error analysis \\
    Free-proton\;PDFs \\
    HQ treatment \\
    Indep.\;flavours \\
    \\
    Reference \\
    \hline
  \end{tabularx}
  \renewcommand{\arraystretch}{0.800}
  \begin{tabularx}{2.4cm}{|>{\columncolor{KSASG20color!10}}Y|}
    \hline
    {\footnotesize \phantom{$^\dag$}KSASG20\phantom{$^\dag$}} \\
    \hline
    \hline
    {\footnotesize NLO \& NNLO} \\
    \checkmark \\
    \checkmark \\
    \checkmark \\
    \\
    \\
    \\
    \\
    \\
    \\
    \\
    \\
    1.3, 0.0 GeV \\
    {\footnotesize N/A} \\
    4353 \\
    9 \\
    {\footnotesize Hessian} \\
    {\footnotesize CT18} \\
    {\footnotesize FONLL} \\
    3 \\
    \\
    \cite{Khanpour:2020zyu} \\
    \hline
  \end{tabularx}
  \hspace{-0.185cm}
  \begin{tabularx}{2.4cm}{|>{\columncolor{TUJU21color!10}}Y|}
    \hline
    {\footnotesize \phantom{$^\dag$}TUJU21\phantom{$^\dag$}} \\
    \hline
    \hline
    {\footnotesize NLO \& NNLO} \\
    \checkmark \\
    \checkmark \\
    \\
    \\
    \\
    \\
    \\
    \\
    \checkmark \\
    \\
    \\
    1.87, 3.5 GeV \\
    {\footnotesize N/A} \\
    2410 \\
    16 \\
    {\footnotesize Hessian} \\
    {\footnotesize own fit} \\
    {\footnotesize FONLL} \\
    4 \\
    \\
    \cite{Helenius:2021tof} \\
    \hline
  \end{tabularx}
  \hspace{-0.185cm}
  \begin{tabularx}{2.4cm}{|>{\columncolor{EPPS21color!10}}Y|}
    \hline
    {\footnotesize \phantom{$^\dag$}EPPS21\phantom{$^\dag$}} \\
    \hline
    \hline
    {\footnotesize NLO} \\
    \checkmark \\
    \checkmark \\
    \checkmark \\
    \checkmark \\
    \checkmark \\
    \\
    \checkmark \\
    {\tiny \phantom{GMVFN}}\checkmark{\tiny GMVFN} \\
    \checkmark \\
    \\
    \\
    1.3, 1.8 GeV \\
    3.0 GeV \\
    2077 \\
    24 \\
    {\footnotesize Hessian} \\
    {\footnotesize CT18A} \\
    {\footnotesize S-ACOT} \\
    6 \\
    \\
    \cite{Eskola:2021nhw} \\
    \hline
  \end{tabularx}
  \hspace{-0.185cm}
  \begin{tabularx}{2.4cm}{|>{\columncolor{nNNPDF30color!10}}Y|}
    \hline
    {\footnotesize \phantom{$^\dag$}nNNPDF3.0\phantom{$^\dag$}} \\
    \hline
    \hline
    {\footnotesize NLO} \\
    \checkmark \\
    \checkmark \\
    \checkmark \\
    \\
    \\
    \\
    \checkmark \\
    {\tiny \phantom{FO+PS}}\checkmark{\tiny FO+PS} \\
    \checkmark \\
    \checkmark \\
    \\
    1.87, 3.5 GeV \\
    0.0 GeV, {\footnotesize N/A} \\
    2188 \\
    256 \\
    {\footnotesize Monte Carlo} \\
    {\footnotesize $\sim$NNPDF4.0} \\
    {\footnotesize FONLL} \\
    6 \\
    \\
    \cite{AbdulKhalek:2022fyi} \\
    \hline
  \end{tabularx}
  \hspace{-0.185cm}
  \begin{tabularx}{2.4cm}{|>{\columncolor{nCTEQ15WZSIHcolor!10}}Y|}
    \hline
    {\footnotesize \phantom{$^\dag$}nCTEQ15HQ$^\dag$} \\
    \hline
    \hline
    {\footnotesize NLO} \\
    \checkmark \\
    \\
    \checkmark \\
    \\
    \checkmark \\
    \checkmark \\
    \\
    {\tiny \phantom{MEfitting}}\checkmark{\tiny ME\,fitting} \\
    \checkmark \\
    \\
    \\
    2.0, 3.5 GeV \\
    3.0 GeV \\
    1484 \\
    19 \\
    {\footnotesize Hessian} \\
    {\footnotesize $\sim$CTEQ6M} \\
    {\footnotesize S-ACOT} \\
    5 \\
    \\
    \cite{Duwentaster:2022kpv} \\
    \hline
  \end{tabularx}
  \vspace{-0.025cm}
  \caption{Recent nPDF global fits.}
  \label{Tbl:nPDFs}
\end{table}
\note[\dag]{See also Refs.~\cite{Segarra:2020gtj,Muzakka:2022wey} for orthogonal developments in the large-$x$ region.}

The nCTEQ15HQ fit uses yet another approach by employing a matrix-element (ME) fitting method in the spirit of Refs.~\cite{Lansberg:2016deg,Kusina:2017gkz,Kusina:2020dki}, including also data on quarkonia production. This works in two steps: first, they fit the open-heavy-flavour and quarkonia matrix elements to proton--proton data (assuming $2 \rightarrow 2$ kinematics and neglecting contributions from initial-state quarks), and second, use the fitted matrix elements to fit the nuclear PDFs with proton--lead data. Though useful in its simplicity, this approach has an obvious downside: the solutions given by the matrix-element fitting are not necessarily those supported by the QCD theory. In particular, the assumption of $g+g$ channel dominance holds only up to a certain precision, and the simple $2 \rightarrow 2$ kinematics get significantly modified by the QCD radiative corrections, as shown in Ref.~\cite{Eskola:2019bgf}. This can, similarly to the FO+PS case, bias the small-$x$ gluon sensitivity of the data to be stronger in the ME-fitting approach than what pQCD would give. These differences in the HQ treatment need to be taken into account when comparing nPDF predictions to data. E.g.\ in Ref.~\cite{LHCb:2022rlh}, the new LHCb measurement of D$^0$ production in pPb at 8.16 TeV is compared only against results from ME fitting in the HELAC formalism~\cite{Lansberg:2016deg,Kusina:2017gkz,Kusina:2020dki}, and therefore still need to be scrutinised with the direct pQCD calculations.

\vspace{-0.275cm}
\section{Gluons in nuclei from heavy to light}
\vspace{-0.175cm}

\begin{figure}
\centerline{%
 \includegraphics[width=0.8\textwidth]{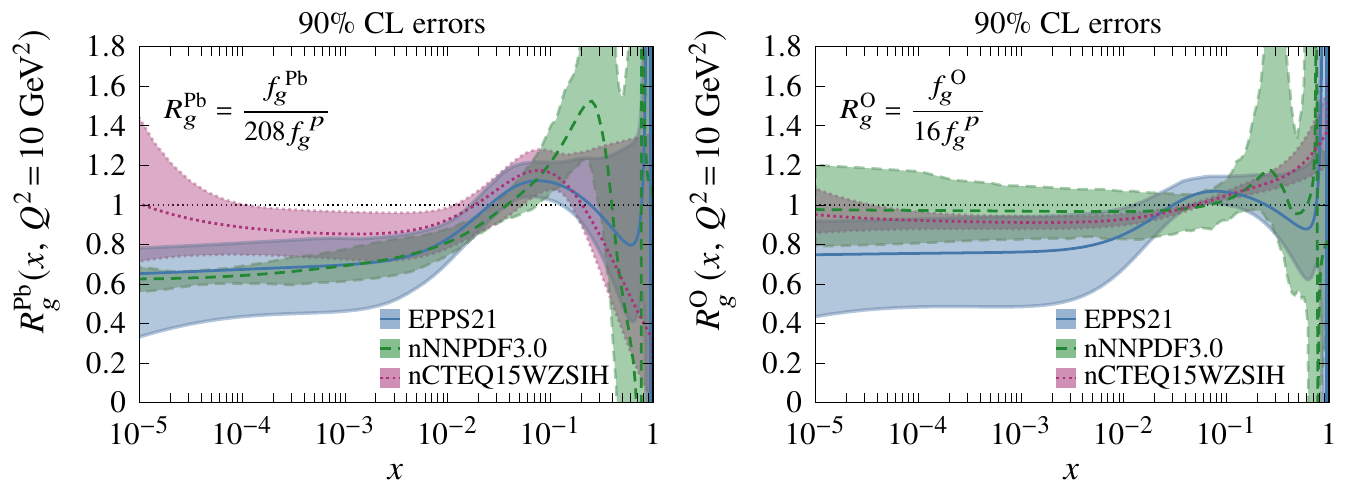}
}
\vspace{-0.125cm}
\caption{Nuclear modifications of the gluon PDFs in lead (left) and oxygen (right) nuclei. Results from the EPPS21~\cite{Eskola:2021nhw}, nNNPDF3.0~\cite{AbdulKhalek:2022fyi} and nCTEQ15WZSIH~\cite{Duwentaster:2021ioo} analyses. See Ref.~\cite{Duwentaster:2022kpv} for the impact of HQ data on the last.}
\label{Fig:comp_nPDFs}
\end{figure}

Figure~\ref{Fig:comp_nPDFs} (left) shows the nuclear modifications in lead nuclei from the EPPS21, nNNPDF3.0 and nCTEQ15WZSIH~\cite{Duwentaster:2021ioo} analyses, the last being the predecessor of nCTEQ15HQ. For EPPS21 and nNNPDF3.0, the gluon constraints are driven by the dijet and D$^0$ data, while in nCTEQ15WZSIH these come largely from EW bosons, preferring smaller shadowing. Notably, the uncertainty bands in EPPS21 and nNNPDF3.0 differ significantly at small $x$, which might be caused by the different treatments of heavy-quark data. In addition, while EPPS21 and nCTEQ15HQ employ a $p_{\rm T}$ cut at 3 GeV, nNNPDF3.0 includes these data all the way to 0 GeV, but also different ways of treating the normalisation uncertainties, the direct fit with look-up tables in EPPS21 versus Monte Carlo reweighting in nNNPDF3.0 and the error-tolerance value in Hessian method used in EPPS21 can all play a role in the size of the small-$x$ uncertainty bands. At large $x$, the gluon modifications in nNNPDF3.0 differ from EPPS21 and nCTEQ15WZSIH, which could be related to the omission of backward D$^0$ and single-inclusive-hadron-production data from nNNPDF3.0. Large differences in the obtained nuclear modifications appear also for lighter nuclei, such as oxygen shown in Figure~\ref{Fig:comp_nPDFs} (right), stemming from the lack of data to directly constrain them. This can cause a major source of uncertainty in testing
small-system energy loss with oxygen-oxygen collisions at the LHC Run~3~\cite{Huss:2020dwe,Brewer:2021tyv}. The need for lighter-than-lead collider-pA data has therefore been recognised~\cite{Paukkunen:2018kmm,Helenius:2019lop,Citron:2018lsq,Paakkinen:2021jjp}.

\vspace{-0.275cm}
\section{Closing remarks}
\vspace{-0.175cm}

The data from past LHC runs have been particularly useful in determining the gluon content in lead nucleus, yielding significant new constraints on the nPDFs. Despite of these advancements, the differences in the extracted gluon PDFs both at small and large $x$ call for new measurements from LHC Run~3. As an example, the increased luminosities might enable the measurement of triple-differential dijets~\cite{Shen:2021eir}, which could put the factorization and nPDFs under a stringent test, and data from proton-oxygen collisions would help determining the gluon PDFs of lighter nuclei~\cite{Paakkinen:2021jjp}.

\vspace{-0.275cm}
\acknowledgments
\vspace{-0.175cm}

The author acknowledges financial support from the Academy of Finland project 330448. This research was funded as a part of the Center of Excellence in Quark Matter of the Academy of Finland and as a part of the European Research Council project ERC-2018-ADG-835105 YoctoLHC.

\bibliographystyle{JHEP}
\bibliography{PoS_LHCP2022_Paakkinen}

\end{document}